# Dual-beam optical linear polarimetry from Southern skies: characterization of CasPol for high-precision polarimetry

Marina Sosa
Carolina von Essen
Ileana Andruchow
Sergio A. Cellone
Luis A. Mammana

SPIE.

# Dual-beam optical linear polarimetry from Southern skies: characterization of CasPol for high-precision polarimetry

**Marina Sosa,**[a,b,*] **Carolina von Essen,**[c] **Ileana Andruchow,**[a,b,d] **Sergio A. Cellone,**[a,d,e] **and Luis A. Mammana**[a,e]
[a]Universidad Nacional de La Plata, Facultad de Ciencias Astronómicas y Geofísicas, La Plata, Argentina
[b]Instituto de Astrofísica de La Plata (CCT-La Plata, CONICET-UNLP), La Plata, Argentina
[c]Aarhus University, Stellar Astrophysics Centre, Department of Physics and Astronomy, Aarhus, Denmark
[d]Consejo Nacional de Investigaciones Científicas y Técnicas, Ciudad Autónoma de Buenos Aires, Argentina
[e]Complejo Astronómico El Leoncito (CONICET-UNLP-UNC-UNSJ), San Juan, Argentina

**Abstract.** We present a characterization of CasPol, a dual-beam polarimeter mounted at the 2.15-m Jorge Sahade Telescope, located at the Complejo Astronómico El Leoncito, Argentina. The telescope is one of the few available meter-sized optical telescopes located in the southern hemisphere hosting a polarimeter. To carry out this work, we collected photopolarimetric data along five observing campaigns, the first one during January 2014, and the remaining ones spread between August 2017 and March 2018. The data were taken through the Johnson–Cousins *V*, *R*, and *I* filters. Along the campaigns, we observed eight unpolarized and four polarized standard stars. Our analysis began characterizing the impact of seeing and aperture into the polarimetric measurements, defining an optimum aperture extraction and setting a clear limit for seeing conditions. Then, we used the unpolarized standard stars to characterize the level of instrumental polarization and to assess the presence of polarization dependent on the position across the charge-coupled device. Polarized standard stars were investigated to quantify the stability of the instrument with wavelength. Specifically, we find that the overall instrumental polarization of CasPol is ∼0.2% in the *V*, *R*, and *I* bands, with a negligible polarization dependence on the position of the stars on the detector. The stability of the half-wave plate retarder is about 0.35 deg, making CasPol comparable to already existing instruments. We also provide measurements in the three photometric bands for both the unpolarized and polarized standard stars. Finally, we show scientific results, illustrating the capabilities of CasPol for precision polarimetry of relatively faint objects. © *2019 Society of Photo-Optical Instrumentation Engineers (SPIE)* [DOI: 10.1117/1.JATIS.5.2.028002]

## 1 Introduction

The first modern polarimetric observations were scheduled to study the reflective properties of the Moon.[1,2] A century later, Chandrasekhar[3] predicted the optical radiation emitted by early-type stars to be polarized. This was the early beginning of a branch of observational astronomy, setting the path to discoveries, such as Serkowski's law.[4] BL Lac objects, a type of active galactic nuclei, have shown high and variable optical linear polarization owing to synchrotron radiation.[5] In these cases, the power of polarimetric measurements relies on the information that they provide on the geometry and orientation of the magnetic field of these sources,[6–8] which is not possible to obtain from photometric data alone.

The southern hemisphere hosts a variety of optical telescopes that can be used to study polarized light of astronomical sources. Examples are the $4 \times 8.2$ m very large telescopes located in Chile. Each one of these telescopes independently hosts several instruments for polarimetric studies, such as NACO[9,10] and SPHERE,[11] both focused in the near-infrared wavelengths. Another example is the Gemini Planet Imager at the 8-m Gemini South telescope, capable of polarimetric imaging at diffraction-limited spatial resolution in the near-infrared.[12] These extremely powerful tools permit researchers to study in detail light from a large variety of sources. However, due to their frontier technology, they are highly oversubscribed, making polarimetric follow-up campaigns of single objects and/or surveys involving large samples unlikely to be scheduled. These observations would then rely on meter-sized telescopes with high-precision instruments. An example of such an instrument located in the southern hemisphere is the 84-cm Robotic Telescope at Cerro Tololo, Chile.[13] Although this instrument can collect high-quality polarimetric data in the optical, it is not suited to follow up intrinsically faint objects that are expected to show photopolarimetric variability, such as blazars, with apparent magnitudes usually fainter than $R \sim 17$.[14]

The Complejo Astronómico El Leoncito (CASLEO), located within the Andes mountains in Argentina, hosts a dual-beam polarimeter, CasPol. This instrument provides one of the few available means to carry out optical photopolarimetric measurements of relatively faint targets from the South, with a telescope that is not heavily oversubscribed. CasPol has been already used to study asteroids[15,16] and is currently used to study the photo-polarimetric microvariability of blazars (Sosa et al., in preparation). These celestial objects of our particular interest are expected to show photopolarimetric variability on the order of a few hours to days, making follow-up campaigns relevant for their study.[17–19] In this work, we present a detailed characterization of the instrumental polarization of CasPol, along with a thoughtful description of the impact of seeing and aperture on our derived polarimetric measurements. From this analysis, we set clear limits to the observing conditions under which precision measurements should be done. In Sec. 3, we present our collected data, our reduction techniques, and a brief description of the construction of the polarimetric data points. In Sec. 4, we describe the impact of seeing and aperture onto the derived polarimetric measurements, we study the instrumental polarization along and across the CCD, and we investigate the dependency of polarization angle with wavelength. Furthermore, we characterize the potential impact of flat-fielding the science frames on the derived polarization values. We then give an illustrative example with results on the blazar 1ES 1101–232, closing in Sec. 5 with discussion and concluding remarks.

---

*Address all correspondence to Marina Sosa, E-mail: marinasosa@fcaglp.unlp.edu.ar

## 2 Generalities about CasPol

The dual-beam polarimeter CasPol follows a design similar to the IAGPOL[20] and the DBIP[21] polarimeters. CasPol is mounted at the Cassegrain focus of the 2.15-m Jorge Sahade telescope. The associated charge-coupled device is a 16 bits CCD TEK of $1024 \times 1024$ pixels, with a plate scale of 0.27 arc sec/pixel. The optical setup provides an unvignetted, circular field-of-view with a diameter of ∼4 arc min. CasPol consists basically of a unit with a mechanical shutter, a filter wheel with UBVRI Johnson filters (the unfiltered option is also available), a neutral filter strip, a half-wave plate (HWP) retarder, and a Savart plate. These two last optical elements have antireflectant coatings between 400 and 800 nm. The HWP can rotate in steps of 22.5 deg, determined by software. Nonetheless, the angle can be changed upon request. The Savart plate produces two orthogonal images of the objects in the field, the so-called ordinary ($O$) and the extraordinary ($E$) beams. These are separated by 0.9 mm, which is equivalent to 10.2 arc sec on the sky. This relatively small separation will constrain the seeing at which polarimetric data should be collected (see Sec. 4.1). Figure 1(a) shows CasPol's field-of-view around the blazar 1ES 1101-232.[22] The original frame has been masked to minimize visual contamination produced by the vignetted region. The corresponding seeing [full width at half maximum (FWHM)] is 2 arc sec and has been estimated from the stars in the field. The figure on the right panel shows an image taken from Aladin Sky Atlas and is placed there as comparison. The black circle indicates approximately coinciding fields.

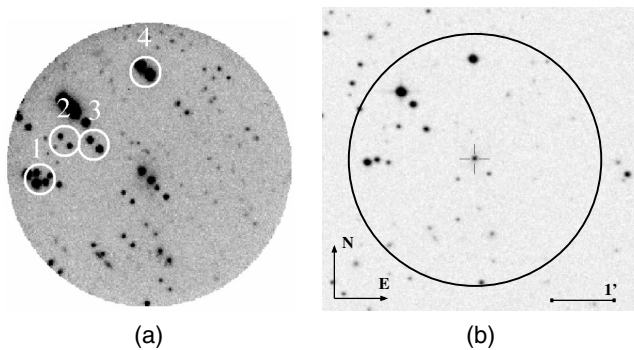

**Fig. 1** Typical field of view of CasPol. (a) The blazar 1ES 1101-232 22 is placed at the centre of the field. The image has been masked to avoid visual contamination by the vignetted area. The double image is caused by the superposition of the O/E beams. The white circle labeled with a number 4 indicates the star used to create Fig. 2. (b) A comparison of the field taken from Aladin. The black circle indicates the approximate coinciding area. As shown in the image, North is up and East is right.

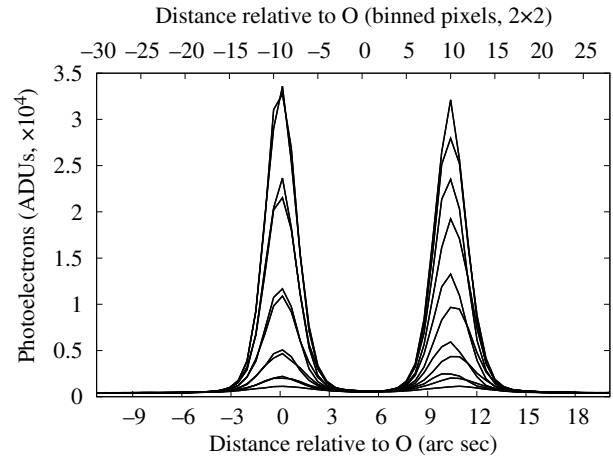

**Fig. 2** Intensity profile of a field star located North from the blazar 1ES 1101-232 (see Fig. 1) in photoelectrons, as shown in arc sec (bottom) and binned pixels (top). Here, the measurements are taken along several diagonal cuts on the detector.

Figure 2 shows the angular separation between the ordinary and extraordinary beams for the star directly North of the blazar [indicated with #4 in Fig. 1(a)], plotted in arcseconds and in $2 \times 2$ binned pixels, as comparison. The figure was performed overplotting several diagonal cuts of the science frame in the vicinity of the mentioned star, thus revealing the maximum count rate and its variation with increasing distance to the $O/E$ centroids.

## 3 Observations and Data Handling

The dual-beam imaging system is very convenient.[21] Through the simultaneous observations of both beams (along with the sky background), photometric conditions can be relaxed, because polarization due to moonlight or dust in our atmosphere is exactly compensated. More importantly, the dual-beam imaging system compensates for photometric variability due to atmospheric turbulence and cancels out unwanted noise caused by passing clouds, water vapor, and aerosols, among others. This observational benefit comes very handy for measurements involving extremely low polarization values, as in the case of polarization of reflected light by exoplanets.[23,24] Here, the polarization levels tend to be of the order of $P \sim 10^{-3}$ to $10^{-4}$. In the particular case of CasPol, due to the small angular separation between the $O/E$ beams, the instrument is mostly suitable for the observation of point sources.

To carry out a thorough characterization of the instrumental polarization of CasPol, we have been granted telescope time along five observing campaigns (OCs) along dark nights exclusively, taking place during January 2014 (OC-1), August 2017 (OC-2), October 2017 (OC-3), January 2018 (OC-4), and March 2018 (OC-5). During these campaigns, we observed 12 polarized and unpolarized standard stars, along with one astronomical source of our particular interest. The standard stars were mostly taken from Ref. 25, which provides a list of standards in good agreement with the visibility during our campaigns. Further standard stars observed and presented in this work can be found under Refs. 26–28. The stars observed during the five campaigns, along with the relevant information about the collected data, can be found in Table 1. The visual apparent magnitudes and the right ascension and declination were extracted from the literature, which is also detailed in Table 1.

**Table 1** From left to right, we detail the name, right ascension ($\alpha$), and declination ($\delta$) in J2000.0, the visual apparent magnitude, $V$, the corresponding bibliographic reference number ($R$ #), the nature (T) of the targets, the observing time (obs. date), the number of nights in which these were observed ($N$), the filter and the collected number of polarimetric points along the five OCs ($F$ # points). P and nP correspond to polarized and nonpolarized standard stars, respectively. The standard stars were taken from Refs. 25 (Ref. #1), 26 (Ref. #2), 27 (Ref. #3), 28 (Ref. #4), 29 (Ref. #5) and 22 (Ref. #6).

| Name | $\alpha$ (J2000.0) | $\delta$ (J2000.0) | $V$ (mag) | R# | T | Obs. Date | N | F (# points) |
|---|---|---|---|---|---|---|---|---|
| HD 10038 | 01:37:18.59 | −40:10:38.5 | 8.14 | 1 | nP | 2017/10/17 | 1 | $V(16), R(15)$ |
| HD 12021 | 01:57:56.14 | −20:05:57.7 | 8.8 | 5 | nP | 2017/08/27 | 1 | $R(2)$ |
| NGC 2024 1 | 05:41:37.85 | −01:54:36.5 | 12.20 | 2 | P | 2018/01/21 | 1 | $V(1), R(1), I(1)$ |
| HD 38393 | 05:44:27.79 | −22:26:54.2 | 3.60 | 4 | nP | 2014/01/28 | 1 | $V(2), R(2)$ |
| HD 42078 | 06:06:41.04 | −42:17:55.7 | 6.16 | 1 | nP | 2018/03/14-17 2018/01/21 2018/10/17 | 6 | $V(10), R(11), I(5)$ |
| HD 64299 | 07:52:25.51 | −23:17:46.8 | 10.01 | 1 | nP | 2018/01/20-21 2014/01/28 | 3 | $V(4), R(4), I(3)$ |
| Ve6 23 | 09:06:00.01 | −47:18:58.2 | 12.12 | 2 | P | 2014/01/28-31 2018/01/20-21 2018/03/14-17 | 10 | $V(10), R(12), I(6)$ |
| HD 298383 | 09:22:29.76 | −52:28:57.4 | 9.75 | 3 | P | 2014/01/29-31 | 3 | $V(4), R(5), I(3)$ |
| HD 94851 | 10:56:44.17 | −20:39:51.6 | 9.29 | 3 | nP | 2018/01/20-22 | 3 | $V(15), R(16), I(15)$ |
| HD 97689 | 11:13:50.75 | −52:51:21.2 | 6.82 | 1 | nP | 2018/03/14-17 | 4 | $V(3), R(3), I(3)$ |
| BD-125133 | 18:40:01.70 | −12:24:06.9 | 10.40 | 1 | P | 2017/08/28 | 1 | $V(2), R(2)$ |
| HD 176425 | 19:02:08.52 | −41:54:37.8 | 6.21 | 1 | nP | 2017/08/28 | 1 | $V(2), R(2)$ |
| 1ES 1101-232 | 11:03:37.61 | −23:29:31.20 | 16.55 | 6 | blazar | 2018/03/16 | 1 | $R(21)$ |

### 3.1 Data Reduction

We performed the data reduction and extraction of the O/E fluxes analyzing all the OCs in a homogeneous way. For this end, we used usual photometric packages of IRAF (CCDPROC/ CCDRED), along with IRAF's scripts created by our research group. All science frames are bias-subtracted. Between campaigns, CasPol was mounted on and dismounted off the telescope. Thus, the position of the shadows of the defocused dust grains that are usually registered by flat-fields changed between campaigns. As a consequence, only when flat-fields taken during a given campaign were available, we also flat-fielded the images. These were taken together with the science frames using the identical optical setup regarding the used filter, the binning, and the angles of the HWP at which the standard stars were observed. In this work, the science frames corresponding to OC-2 and OC-3 are not calibrated by flats, while the ones corresponding to OC-1, OC-4, and OC-5 are (see Sec. 4.5 for a detailed analysis) calibrated. We computed photometric fluxes using our own IRAF task, MULTIFOT. The task runs phot interactively and is suitable to automatically extract O/E fluxes from the science frames. We integrated fluxes in several apertures to investigate the impact of its choice on the derived polarimetric values (see Sec. 4.1 for an analysis in more detail).

### 3.2 Construction of Polarimetric Points

To construct the Stokes $Q$ and $U$ parameters, we followed Ref. 30 and computed the intermediate values:

$$R_Q^2 = \frac{I_0^O/I_0^E}{I_{45}^O/I_{45}^E}, \quad R_U^2 = \frac{I_{22.5}^O/I_{22.5}^E}{I_{67.5}^O/I_{67.5}^E}, \quad (1)$$

taking into account that one polarimetric point was observed rotating the HWP retarder to angles of 0 deg, 22.5 deg, 45 deg, and 67.5 deg. In general, $I_\beta^O$ and $I_\beta^E$ are the object ordinary and extraordinary integrated fluxes, respectively, and $\beta$ is the position angle of the HWP.[18,31] The Stokes parameters are then computed from these values as follows:

$$Q = \frac{R_Q - 1}{R_Q + 1}, \quad U = \frac{R_U - 1}{R_U + 1}. \quad (2)$$

Based on these parameters, we calculated the degree of linear polarization and corresponding polarization angle in the usual way:

$$P = \sqrt{Q^2 + U^2}, \quad \Theta = \frac{1}{2}\arctan\left(\frac{U}{Q}\right). \quad (3)$$

Error estimates for the Stokes $Q$ and $U$ parameters, the polarization degree, and the polarization angle are computed following standard error propagation techniques. These, in turn, depend on the uncertainties on the fluxes given by IRAF's PHOT. Here, the error associated to a flux measurement is based on three terms. These are the photon noise within the aperture, the standard deviation of the pixels comprising the sky ring that are used to determine the background, and a term that accounts for the uncertainty in the background level.[32] Our derived error estimates were verified with and compared to the ones available in Ref. 33. Our uncertainties for the polarization degree, $\sigma_P$, and the polarization angle, $\sigma_\Theta$, are as follows:

$$\sigma_P = (Q^2\sigma_Q^2 + U^2\sigma_U^2)^{1/2}\frac{1}{P},$$
$$\sigma_\Theta = (Q^2\sigma_U^2 + U^2\sigma_Q^2)^{1/2}\frac{1}{2P^2}. \quad (4)$$

As pointed out by Ref. 32, it is known that the photometric errors determined by IRAF are underestimated and, in consequence, individual errors on $P$ and $\Theta$ are underestimated as well. However, it is worth to mention that when computing polarimetric measurements from a set of points, we always computed errors in two ways. These are from error propagation and computing the standard error of the mean for objects assumed to be nonvariable. The latter uses the natural scatter of the data and, thus, reflects more realistically the precision of our measurements.

Since polarization is positive-definite, when calculating the polarization degree, the noise in their involved quantities contribute in a positive way, producing biased results. This has been addressed by numerous authors,[34–36] also detailed in Ref. 19. To correct for this bias, for all the unpolarized standard stars, we computed the unbiased degree of linear polarization, $P_{unbiased}$, using the expression found in Ref. 35:

$$P_{unbiased} = \sqrt{P^2 - a \times \sigma P^2}. \quad (5)$$

Here, $P_{unbiased}$ was computed using the maximum likelihood estimator that can be found in their work ($a = 1.41$) and $\sigma P = (\sigma Q + \sigma U)/2$. To determine when it is necessary to apply the bias correction, we followed the selection criteria adopted and described in Ref. 36. The authors assume that a given celestial object is polarized if the lower confidence limit (95%) of $P$ is $>0$. In this case, $P_{unbiased}$ is obtained computing Eq. (5). If $P$ is consistent with 0, then an upper limit is assigned to $P$ considering the upper 95% confidence limit. The corrected values of polarization, $P_{unbiased}$, are listed in Tables 2 and 3, and shown in Figs. 7–9.

## 4 Results

### 4.1 Testing the Impact of Seeing and Aperture on Our Polarimetric Measurements

The data collected during the five OCs comprise different observing conditions, mostly reflected as changes of airmass and seeing during observations. In order to compare the photo-polarimetric values derived from these data, it is fundamental to find an extraction aperture common to all the campaigns that both minimizes the scatter of the polarimetric measurements and maximizes the signal-to-noise ratio (SNR) of the individual points. As shown in previous sections, the separation between ordinary and extraordinary beams is fixed to 10.2 arc sec (equivalently, ∼38 unbinned pixels). To avoid the inclusion of a significant amount of flux from the extraordinary beam within the aperture centered on the ordinary image (and vice versa), one half of this separation should not be exceeded.

**Table 2** Derived values for the instrumental polarization of CasPol as a function of the photometric band. Uncertainties are given at $1\sigma$ level.

| Band | $Q$ (%) | $U$ (%) | $P$ (%) | $P_{unbiased}$ (%) |
|---|---|---|---|---|
| V | −0.12 ± 0.03 | 0.10 ± 0.05 | 0.16 ± 0.03 | 0.15 ± 0.03 |
| R | −0.10 ± 0.06 | 0.03 ± 0.06 | 0.10 ± 0.06 | <0.22 |
| I | −0.03 ± 0.08 | 0.09 ± 0.04 | 0.09 ± 0.05 | <0.19 |

To sample the effects of aperture and seeing adequately, we measured O/E fluxes for all the unpolarized standard stars with circular aperture radii ranging from ∼0.5 to ∼7 arc sec with steps of ∼0.5 arc sec (equivalently, from 2 to 26 pixels, each 2 pixels). For all the derived polarimetric measurements, we observe a similar behavior, but for a better and clear visualization, we only showed the results of some representative targets in Fig. 3. For apertures lower than the mean FWHM of a given polarimetric point, the derived polarimetric values and their scatter are large and inconsistent with zero. Due to the rapid changes in seeing during observations, the shape of the point-spread functions (PSFs) suffers irregular deformations, which differ between the O/E beams. These deformations are enhanced at the core of the PSFs. Thus, when integrating within small apertures (i.e., mostly the cores at typical values of seeing of 2 to 3 arc sec), these differences translate into the systematic increase of polarimetric scatter. For larger apertures, particularly larger than half the angular separation between O/E beams, we observe again an increment in the polarimetric scatter, but not as large as for apertures smaller than the FWHM. This effect is caused because the wings of the extraordinary beam contribute significantly to the flux within the photometric aperture of the ordinary image (and vice versa), and it can be especially noted during observing nights with poor seeing. Taking these aspects into consideration, for nights when the seeing was low (∼2 arc sec or lower), aperture sizes can reliably range from 3.5 up to 5 arc sec. If seeing is large (typically 3 to 4 arc sec) or significantly variable during an observing run, it is convenient to take larger apertures[37] but always smaller than half the separation between the O/E centroids. To consider these aspects simultaneously, throughout this work, fluxes are integrated using apertures of 5 arc sec. This value was obtained by fitting a second-order polynomial—through a simple least-squares minimization technique—to the aperture-dependent polarimetric points (the minimum of the polynomial is exactly at 4.67 arc sec) and to their scatter (4.95 arc sec, respectively). The final value of 5 arc sec conservatively considers these two aspects. It is worth to mention that observations taken with seeing larger than ∼6 arc sec should not be used for scientific purposes, because of the contamination between the O/E beams. On the other hand, seeing values between ∼2 and ∼5 arc sec should not be necessarily thought as bad, especially if seeing is constant along the night. In sparse fields, the natural defocusing that these seeing values produce can significantly improve the photometric precision of CCD data,[38] because the noise associated to the intrapixel response variability can be better averaged out when the PSFs are spread over many pixels. This is particularly relevant for telescopes without guiding system and can boost the photopolarimetric precision of astronomical data.

**Table 3** Stokes $Q$ and $U$ values, and polarization degree for the unpolarized standard stars observed during OC-1 to OC-5, corrected by instrumental polarization and rotated to the standard system. Errors are at $1\sigma$ level. When available, the last column shows published values for the polarization degree.

| Name | Filter | $Q_{CasPol}$ (%) | $U_{CasPol}$ (%) | $P_{CasPol}$ (%) | $P_{CasPol}^{unbiased}$ (%) | $P_{pub}$ (%) |
|---|---|---|---|---|---|---|
| HD 42078 | V | 0.10 ± 0.03 | 0.04 ± 0.09 | 0.12 ± 0.06 | 0.09 ± 0.07 | 0.07 ± 0.01 |
|  | R | 0.13 ± 0.1 | −0.08 ± 0.1 | 0.19 ± 0.1 | <0.39 | … |
|  | I | 0.08 ± 0.07 | 0.03 ± 0.1 | 0.09 ± 0.08 | <0.25 | … |
| HD 97689 | V | 0.05 ± 0.02 | −0.11 ± 0.02 | 0.12 ± 0.02 | 0.12 ± 0.02 | 0.14 ± 0.08 |
|  | R | 0.00 ± 0.05 | −0.04 ± 0.08 | 0.04 ± 0.08 | <0.2 | … |
|  | I | −0.04 ± 0.2 | 0.02 ± 0.1 | 0.04 ± 0.2 | <0.44 | … |
| HD 176425 | V | 0.18 ± 0.04 | −0.09 ± 0.01 | 0.20 ± 0.04 | 0.19 ± 0.04 | 0.17 ± 0.03 |
|  | R | 0.17 ± 0.02 | −0.04 ± 0.02 | 0.18 ± 0.02 | 0.18 ± 0.02 | … |
|  | I | … | … | … | … | … |
| HD 94851 | V | 0.15 ± 0.04 | −0.03 ± 0.02 | 0.15 ± 0.04 | 0.14 ± 0.04 | … |
|  | R | 0.02 ± 0.02 | 0.00 ± 0.01 | 0.02 ± 0.02 | <0.06 | … |
|  | I | −0.10 ± 0.04 | 0.02 ± 0.02 | 0.10 ± 0.04 | 0.09 ± 0.04 | … |
| HD 10038 | V | 0.12 ± 0.01 | −0.15 ± 0.02 | 0.19 ± 0.01 | 0.19 ± 0.01 | 0.11 ± 0.01 |
|  | R | 0.01 ± 0.07 | −0.04 ± 0.08 | 0.05 ± 0.07 | <0.19 | … |
|  | I | … | … | … | … | … |
| HD 12021 | V | … | … | … | … | 0.078 ± 0.018 |
|  | R | −0.00 ± 0.43 | −0.10 ± 0.2 | 0.10 ± 0.1 | < 0.3 | … |
|  | I | … | … | … | … | … |
| HD 38393 | V | 0.12 ± 0.07 | −0.11 ± 0.07 | 0.16 ± 0.07 | 0.14 ± 0.08 | 0.0006 ± 0.0003 |
|  | R | 0.09 ± 0.07 | −0.19 ± 0.09 | 0.21 ± 0.07 | 0.19 ± 0.08 | … |
|  | I | … | … | … | … | … |
| HD 64299 | V | 0.06 ± 0.02 | −0.06 ± 0.06 | 0.08 ± 0.05 | <0.18 | 0.06 ± 0.07 |
|  | R | −0.02 ± 0.03 | −0.09 ± 0.03 | 0.09 ± 0.03 | 0.08 ± 0.03 | … |
|  | I | −0.13 ± 0.01 | −0.09 ± 0.00 | 0.16 ± 0.01 | 0.16 ± 0.01 | … |

## 4.2 Instrumental Polarization

All polarimeters have sources that can introduce instrumental polarization that need to be carefully removed to faithfully recover the true polarization, specifically, design factors and optical setup. Possibly, in CasPol, the main contribution arises within the telescope mirrors.[39] To characterize the level of instrumental polarization introduced by CasPol, we focused our analysis on the in-depth study of three unpolarized standard stars that were observed during OC-2, OC-3, and OC-5, namely HD 42078, HD 97689, and HD 176425. Even though our sample of unpolarized standard stars is larger than these three, some of them present additional challenges that we consciously wanted to avoid when characterizing the level of instrumental polarization of CasPol. For instance, HD 64299 has been initially catalogued as being an unpolarized standard star.[27] However, in a subsequent work,[21] a 0.1% polarization level was detected. In addition, HD 94851 and HD 10038 were not included in this analysis because they were observed purposely in different positions across the CCD. These observations were used to characterize the instrumental polarization dependent on position. To minimize contamination introduced by instrumental artifacts, such as potential polarization dependent on position, the three unpolarized standard stars were placed at the exact same positions, coinciding with the center of the CCD.

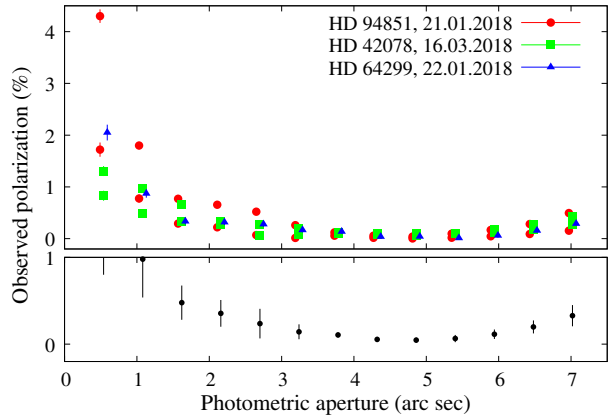

**Fig. 3** Top: values of polarization degree in percentage for three unpolarized standard stars as a function of photometric aperture size, in arc sec. Different minor horizontal offsets, where applied for a better visualisation. Bottom: their respective means (dots) and standard deviations (bars), focused on polarization degrees smaller than 1%. These were obtained averaging all polarimetric measurements per aperture.

To compute the averaged Stokes $Q$ and $U$ parameters, we used the data corresponding to the three unpolarized standard stars collected along the three OCs. Before doing so, we visually inspected the Stokes $Q - U$ values to identify and discard outliers. To do so, we made use of the generalized extreme studentized deviate (ESD) test.[40] The test can be used for a dataset, which follows approximately a normal distribution. In our case, a given point was identified as outlier if the distances to its right and left neighbors are abnormal as judged by the generalized ESD. As a conservative distance, we used five times the standard deviation of the points. A posterior individual checkup of the outliers resulted in corresponding poor photometric signal. Then, for each one of the stars, we computed the differences between the averaged $(Q, U)$ values and the ones tabulated in the literature, and averaged these differences among the three stars to arrive to the final instrumental polarization. The three unpolarized standard stars do not have published values for both $R$ and $I$. However, between 2010 and 2016[26] observed these stars in the mentioned filters. Rather than reporting final values for their polarization state, the authors provide the individual values along with an estimate of the SNR of their measurements. Thus, as reference, we used the values that show the largest SNR. Errors were computed in two ways, from error propagation and computing the standard error of the mean, $\sigma/\sqrt{n}$. Here, $\sigma$ corresponds to the standard deviation of the Stokes parameters and $n$ corresponds to the number of polarimetric points. To be as conservative as possible, we chose as final error the largest one of these two. For each photometric band, we repeated the same procedure. Table 2 shows our derived values for the instrumental polarization of CasPol, for the $V$, $R$, and $I$ filters. We find the level of instrumental polarization of CasPol to be lower than ~0.2%.

### 4.3 Polarized Standard Stars

To compare our derived polarimetric measurements with values from the literature, after correcting for instrumental polarization,

it is required to convert our measurements to the standard system. To do so, we observed four polarized standard stars during our OCs. To determine the adequate rotation angle in order to rotate the data to the standard system, we analyzed the polarimetric data of only three of them, namely NGC 2024 1, Ve6-23, and HD 298383. The procedure was similar to the one carried out to analyze the unpolarized standard stars. From the averaged Stokes $Q$ and $U$ values, we computed the observed polarization angle and the correction angle in the following way: $\Delta\theta = \theta_{\text{bib}} - \langle\theta_{\text{obs}}\rangle$. Here, $\theta_{\text{bib}}$ corresponds to the published polarization angle of the polarized standard stars, while $\langle\theta_{\text{obs}}\rangle$ corresponds to the observed polarimetric angle determined from our averaged $Q$ and $U$ values. We carried out this procedure for each polarized standard star and each photometric band. Afterward, we computed a correction per filter, averaging the individual corrections determined from each one of the three stars. The derived values are $\Delta\theta_V = -4.2 \pm 0.2$ deg, $\Delta\theta_R = -4.3 \pm 0.6$ deg, and $\Delta\theta_I = -4.2 \pm 0.7$ deg.

After characterizing the instrumental polarization of CasPol and determining the $\Delta\theta$ that allows us to report values in the standard system, we corrected all the remaining unpolarized and polarized standard stars by both effects, always in the Stokes $Q - U$ plane. The resulting values are listed in Table 3 for the unpolarized stars and Table 4 for the polarized stars. In most cases, we can assess the goodness of our procedure by comparing the polarization level between CasPol and the values reported in the literature. In the remaining cases, there were no reported values for polarization in all (or some) of our three observed bands. In consequence, for several unpolarized and polarized standard stars, we also report, for the first time, their wavelength-dependent polarization degree and angle. This holds right for HD 12021, HD 38393, HD 64299, and HD 298383. In almost all cases, our derived quantities for the polarized stars in the $V$ band are in agreement at $1\sigma$ level with published values. In the case of the unpolarized stars, this percentage is about 50%.

A usual way to test the stability of a polarimeter in the different wavelengths is by analyzing the behavior of polarization as a function of color.[26] The polarization generated by interstellar dust is a component associated with the Galaxy, generated by the orientation of interstellar dust particles with respect to the magnetic field of the Milky Way.[41] Since we are sampling different grain populations with different sizes, composition and shapes for a constant position angle, we expect to observe a constant change rate between wavelength and polarization. To test the stability of CasPol, we observed the polarized standard stars in the Johnson–Cousins $V$, $R$, and $I$ filters. Figure 4 shows the derived values for the polarization degree, the Stokes $Q$ and $U$ parameters, and the polarization angle for Ve6 23. The same polarized standard star was observed and analyzed by Ref. 26 (Vela 1 95), allowing for a comparison between results. As a main difference with Ref. 26, data present a continuous wavelength coverage between 4000 to 9500 Å, while our observations comprise only three broad-band photometric filters. Nonetheless, to compare our results with Ref. 26, we fitted to our polarimetric values Serkowski's law:[4]

$$p(\lambda)/p_{\max} = \exp[-K \ln^2(\lambda_{\max}/\lambda)], \qquad (6)$$

where the fitting parameters, $p_{\max}$ and $K$, correspond to the peak polarization level and the width constant, respectively. Due to the discontinuous nature of our data, we used the value $\lambda_{\max} = 5864$ Å reported by Ref. 26. To obtain the best fit values and their errors, we sampled from the posterior probability distributions using a Markov-chain Monte Carlo approach, all wrapped up in Python routines that make use of the PyAstronomy[42] package. In this work, errors are given as 68.3% highest probability density credibility intervals. Our derived values are $p_{max} = 8.17 \pm 0.05\%$ and $K = 0.96 \pm 0.10$, inconsistent to the ones reported by Ref. 26. We noted, however, that the authors also found an inconsistency with the values reported by Ref. 43. Changing $\lambda_{max}$ to 5606 Å,[43] and refitting the polarization level and the width constant afterward, results in $p_{max} = 8.08 \pm 0.03\%$, and $K = 1.28 \pm 0.03$. These values are in full agreement at $1\sigma$ level with.[43] The reasons behind this incompatibility escape the scope of this paper; however, we believe they can be related to different transmission functions of the $I$-band filters, combined with different quantum efficiency drops of the CCDs, that strongly diverge around the $I$ wavelengths.

**Table 4** Derived measurements of the Stokes parameters $Q$, $U$, the polarization degree and the polarization angle for the polarized standard stars observed between OC-1 and OC-5. As comparison, published values of polarization degree and angle, when available.

| Name | Filter | $Q_{CasPol}$ (%) | $U_{CasPol}$ (%) | $P_{CasPol}$ (%) | $\theta_{CasPol}$ (deg) | $P_{pub}$ (%) | $\theta_{pub}$ (deg) |
|---|---|---|---|---|---|---|---|
| NGC 2024 1 | V | 0.25 ± 0.09 | −9.9 ± 0.09 | 9.92 ± 0.09 | 135.7 ± 0.3 | 9.548 ± 0.013 | 135.94 ± 0.02 |
|  | R | 0.09 ± 0.09 | −9.6 ± 0.09 | 9.6 ± 0.09 | 135.3 ± 0.3 | 9.671 ± 0.004 | 135.93 ± 0.01 |
|  | I | 0.01 ± 0.09 | −8.4 ± 0.09 | 8.41 ± 0.09 | 135.0 ± 0.3 | 9.009 ± 0.002 | 135.90 ± 0.01 |
| Ve6-23 | V | 7.85 ± 0.03 | −2.17 ± 0.03 | 8.14 ± 0.03 | 172.3 ± 0.1 | 8.163 ± 0.011 | 172.41 ± 0.02 |
|  | R | 7.51 ± 0.03 | −2.14 ± 0.02 | 7.81 ± 0.03 | 172.0 ± 0.1 | 7.927 ± 0.003 | 172.06 ± 0.01 |
|  | I | 6.00 ± 0.03 | −1.69 ± 0.04 | 6.23 ± 0.03 | 172.1 ± 0.2 | 7.151 ± 0.002 | 171.95 ± 0.01 |
| HD 298383 | V | 2.4 ± 0.09 | −4.62 ± 0.08 | 5.19 ± 0.08 | 148.6 ± 0.5 | 5.23 ± 0.09 | 148.6 |
|  | R | 2.49 ± 0.05 | −4.68 ± 0.07 | 5.30 ± 0.07 | 149.0 ± 0.3 | — | — |
|  | I | 1.88 ± 0.05 | −3.72 ± 0.02 | 4.17 ± 0.03 | 148.4 ± 0.3 | — | — |
| BD 125133 | V | 2.0 ± 0.09 | −3.84 ± 0.08 | 4.31 ± 0.09 | 148.6 ± 0.9 | 4.37 ± 0.04 | 146.84 ± 0.28 |
|  | R | 1.42 ± 0.02 | −3.65 ± 0.06 | 3.92 ± 0.06 | 145.6 ± 0.2 | 4.02 ± 0.02 | 146.97 ± 0.13 |
|  | I | — | — | — | — | 3.57 ± 0.09 | 143.99 ± 2.27 |

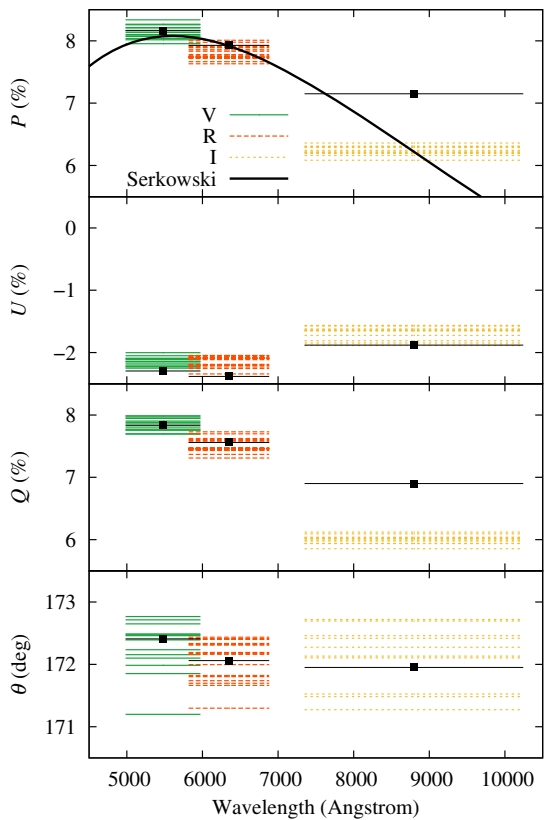

**Fig. 4** From top to bottom: values of the polarization degree and the Stokes $U$ and $Q$ parameters, in percentage, and the angle, in deg, for the polarized standard star Ve6 23. The derived values are shown per photometric band. The widths of the lines correspond to the FWHM of the filter responses. The black line shows our best-fit Serkowski law. Black squares with error bars correspond to the values reported by Ref. 26.

Figure 5 shows our derived $Q$, $U$ values corrected for instrumental polarization for the three photometric bands and rotated to the standard system. The overplotted black lines limit the regions of the measurements derived by Ref. 26. To quantify the change in polarization with wavelength, we fitted to these data points a first order, wavelength-dependent polynomial, $f(\lambda) = a\lambda + b$, with parameters $a = -0.26 \pm 0.02$ and $b = -0.12 \pm 0.17$. The derived slope is consistent with the one observed by Ref. 26.

The polarized standard stars also allow us to quantify the stability of the HWP retarder.[26] Figure 6 shows the residuals of the mean $\theta$ values for all our polarized standard stars that were in turn computed subtracting to each polarization angle the corresponding mean value of the polarization angle. The points are color-coded in the usual way according to the photometric band, and different symbols correspond to different stars. The data points are plotted as a function of time in Julian dates and have been arbitrarily shifted to allow for a better visual inspection of the individual campaigns. The four observing groups

Sosa et al.: Dual-beam optical linear polarimetry from Southern skies...

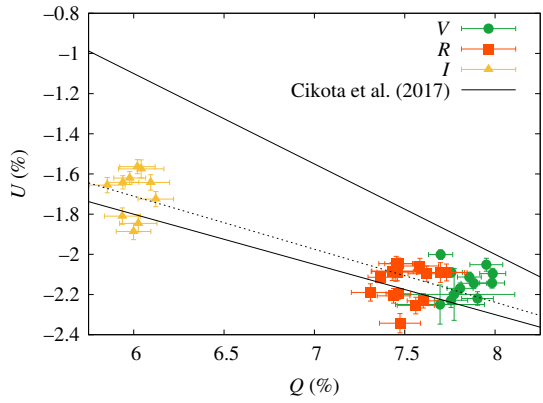

**Fig. 5** $Q/U$ diagram for Ve6 23 in percentage, as a function of wavelength. The black lines show the approximate range of (Ref. 26) $Q/U$ points.

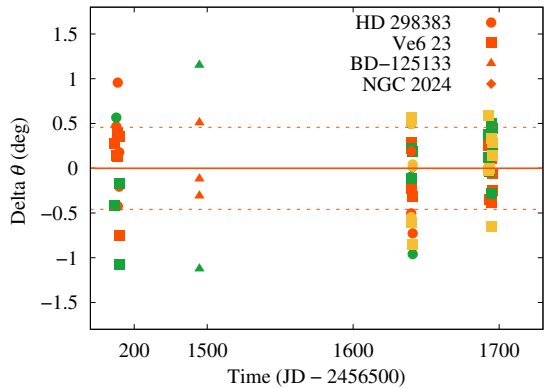

**Fig. 6** Stability of the polarization angle of CasPol as a function of time in Julian dates (JD) along four OCs. Our $V$, $R$, and $I$ measurements are color-coded in the usual way. Different symbols correspond to different polarized standard stars.

where polarized standard stars were observed can be clearly identified. We determined the stability of the HWP from the standard deviation of the residuals of the angles. We do not observe any systematic effect with wavelength. While the variability amplitudes are $\text{Amp}_V = 2.2$ deg, $\text{Amp}_R = 1.7$ deg, and $\text{Amp}_I = 1.4$ deg, the standard deviations are $\sigma_V = 0.55$ deg, $\sigma_R = 0.39$ deg, $\sigma_I = 0.45$ deg. Note, however, how the scatter of the residuals of the angles decreases toward the last two OCs. During the first and second campaigns, the seeing was high, almost reaching our imposed limit of 6 arc sec. Neglecting these campaigns and limiting our stability analysis to the last two, the averaged amplitude of variability is 1.8 deg, while the standard deviation of the residuals is $\Delta\theta = 0.35$ deg, averaged over all the bands.

### 4.4 Observed Polarization with Position on the CCD

To quantify the behavior of instrumental polarization with position on the CCD, following Ref. 33, we observed unpolarized standard stars in locations as homogeneously distributed as possible on the CCD. Here, we focus our analysis on the unpolarized standard stars HD 94851 and HD 10038. While the first one was observed in 12 different positions and in the three photometric bands, the second one was observed in 15 positions but only in the $V$ and $R$ bands. The measured polarimetric values for HD 94851, as a function of the $(X, Y)$ positions that correspond to the centroid of the O image, are shown in Fig. 7 exemplifying our results. The science frames were taken in the binning $2 \times 2$ configuration. In particular, the $(X, Y)$ positions are derived averaging the $(X, Y)$ co-ordinates of the centroids of the four images used to construct each polarimetric point, to account as much as possible for irregularities in the tracking of the telescope. The figures show a square contained within the unvignetted area of the CCD, where we placed the stars. The base of the arrows indicates the exact locations of the stars. The maps are a bilinear interpolation of the sampled points.

To characterize the dependence of polarization with position of the star on the CCD, we computed the Pearson[44] and the Spearman[45] correlation coefficients between the polarization degree and the $X$ position, the $Y$ position, and the distance $d$ with respect to the center of the CCD, following the pre-established tasks in the Python-SciPy library. To be conservative, the values listed in the second and fourth column of Table 5 always correspond to the largest value of the two correlation coefficients. The derived values do not show a strong correlation for the $V$ and $R$ bands, and only marginal for the $I$ band. Despite the magnitude of the derived values, it is important to evaluate the confidence interval of the correlation coefficients. For this end, we made use of the bootstrapping technique. Here, we kept the same polarization values fixed, but we randomly permuted their corresponding $X$, $Y$, and $d$ values $10^4$ times each. For each bootstrapped sample, we computed the Pearson/Spearman correlation coefficient between the polarization degree and the

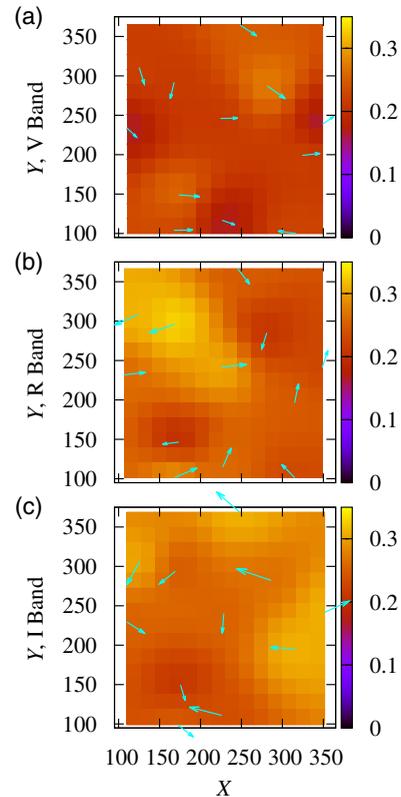

**Fig. 7** (a)–(c) Map of polarization degree in percentage as a function of the X and Y positions of the O beams on the CCD for the $V$, $R$, and $I$ bands, respectively. Cyan arrows show the directions of their respective Stokes $Q$, $U$ parameters corrected for instrumental polarization and rotated to the standard system; they have been equally enlarged to meet the scale of the figures.

**Table 5** Correlation coefficients for P versus position on the CCD, corresponding to both unpolarized standard stars in V, R, and I. Values for the original data points and the shuffled ones (subindex S) are given (see text). P corresponds to the polarization degree, X and Y to the CCD positions in pixels, and d to the distance from the centre of the CCD.

|     | HD 94851 | HD94851$_S$ (%) | HD 10038 | HD10038$_S$ (%) |
|-----|----------|-----------------|----------|-----------------|
|     |          | V               |          |                 |
| P/X | 0.06     | 90              | 0.22     | 50              |
| P/Y | 0.07     | 80              | −0.20    | 49              |
| P/d | 0.23     | 48              | −0.18    | 48              |
|     |          | R               |          |                 |
| P/X | −0.35    | 26              | −0.03    | 87              |
| P/Y | 0.24     | 43              | 0.35     | 22              |
| P/d | −0.18    | 57              | −0.11    | 75              |
|     |          | I               |          |                 |
| P/X | 0.35     | 21              | …        | …               |
| P/Y | 0.46     | 10              | …        | …               |
| P/d | 0.38     | 21              | …        | …               |

permuted X, Y, and d values. Then, we simply counted the number of times that the derived correlation was larger than the original one. After the total iterations were reached, we computed the percentage of exceeding the correlation coefficient as the number of times the correlations from the shuffled values were larger than the real one, divided by the total number of iterations. The strength of the correlation value is measured inversely to the derived percentage. Thus, a large percentage implies a weak correlation signal. In addition, from the $10^4$ correlation values computed for X, Y, and d, we determined their mean and standard deviations. In all cases, the mean values for the correlations were close to 0, while the standard deviations were of the order of 0.3. Thus, we interpret all correlation values between ±0.3 to be inconsequential.

To quantify stability in the polarimetric maps, we computed two variability indices, namely the ratio of the standard deviation to the sample mean as follows:

$$\frac{\sigma}{\mu} = \frac{\sqrt{\sum_{n=1}^{N}(x_n - \mu)^2/(N-1)}}{\sum_{n=1}^{N} x_n / N}, \quad (7)$$

where N is the total number of polarimetric points, and the ratio of the mean square successive difference to the variance of the polarimetric points is as follows:[46]

$$\eta = \frac{\delta^2}{\sigma^2} = \frac{\sqrt{\sum_{n=1}^{N-1}(x_{n+1} - x_n)^2/(N-1)}}{\sigma^2}. \quad (8)$$

For the ratio of the standard deviation to the sample mean, a large quotient ($\frac{\sigma}{\mu} > 2$) implies strong variability. In all photometric bands, the derived ratio is well below 0.5, thus no significant variability was detected in the polarization degree. In the second case, if serial correlation exists (i.e, the relationship between a given point and a lagged version of itself over various time intervals), the ratio is significantly high or small. Our derived values range between $\eta = 2.2$ and $\eta = 2.6$, thus showing no serial correlation between consecutive polarimetric points in neither of the photometric bands. This is illustrated in Fig. 7, where visual inspection does not reveal any correlation pattern between polarization and position, as reported by, for instance,[36] in the CAFOS polarimeter.

Besides the magnitude of the polarimetric points, the direction of the position-dependent polarization values could be suffering from instrumental artifacts. To check if the derived $(Q, U)$ values are randomly distributed on the Stokes $Q - U$ plane or do show any systematic trend, we carried out the following exercise (for both unpolarized stars, in each photometric band). First, we shifted the $(Q, U)$ values around $(Q, U) = (0,0)$ subtracting to each $(Q, U)$ pair their respective averages. Afterward, we counted how many $(Q, U)$ pairs were placed in each quadrant, $N_{\text{quad}}$, and kept these four numbers for future analyses. From here, the quadrants to which we refer are: +Q+U, +Q−U, −Q−U, and −Q+U. After this, we generated fake $(Q, U)$ values that were randomly distributed about the four quadrants. The length of this fake set of pairs equals the length of the real data. Once generated, we counted how many $(Q, U)$ pairs were placed in each quadrant, $N_{\text{quad,fake}}$, and kept this number as reference.

Knowing that the nature of the fake $(Q, U)$ values is random, we wanted to quantify how many times randomly generated $(Q, U)$ values would fall in the four quadrants, equally, when compared to the observed Stokes parameters. To quantify this number, we iterated $10^5$ times. At each iteration, we computed a set of $(Q, U)$ values randomly distributed with the same size as the real sample and counted how many times the number of elements per quadrant was the same as $N_{\text{quad,fake}}$. This percentage, with typical values around 3%, was used as reference. We repeated this exact same exercise but using the real $(Q, U)$ values. All this process was repeated $10^4$ times. If $N_{\text{quad}}$ is at least as large as $N_{\text{quad,fake}}$, then we understand the nature of the $(Q, U)$ values to be as random as the fake data, i.e., random. Counting the number of times that $N_{\text{quad}} > N_{\text{quad,fake}}$, divided by the total number of iterations, gives us an idea of the strength of the randomness of the $(Q, U)$ values. For the different photometric bands, the derived percentages are $\text{Rand}_V = 75\%$, $\text{Rand}_R = 57\%$, and $\text{Rand}_I = 60\%$. As an internal checkup of this procedure, we consciously used the absolute value of the Stokes $Q$, $U$ parameters instead the real values, shifting them in this way to the first quadrant exclusively. The derived percentages are, as expected, significantly lower. The corresponding values are $\text{Rand}_V = 14\%$, $\text{Rand}_R = 15\%$, and $\text{Rand}_I = 12\%$.

### 4.5 Flat-fielding

An important matter to define before analyzing polarimetric data taken with a dual-beam imaging polarimeter is what kind of flats are going to be used to calibrate the science frames. Ideally, the light collected by flat frames should suffer the same effects than stellar light. However, the sources of light used to acquire flats are not homogeneously illuminated (either sky or dome flats). To minimize this effect, a technique that can be used is to create a master flat averaging individual flats taken in all the angles at which the HWP was rotated.[47] However, this procedure is not always effective because the intensity of the source of light used to acquire the flats is not really stable.[33] An alternative, carried out in this work, and also chosen by some groups[48,49] is to take a minimum of 10 flats for each angle of the HWP, creating afterward a master flat per rotation angle. Then, each science frame should be divided depending on the angle of the HWP at which it was taken. We also used unflattened science frames, as a test. The derived $Q$ and $U$ parameters of the unpolarized standard stars were divided into two groups: those obtained from flat-fielded images (62 polarimetric points) and those corresponding to nights with no flat-fields available (42 points). After verifying that the $Q$, $U$ points are normally distributed, we compared the two samples by carrying out a Z-test.[50] Here, the null hypothesis is that the two set of points are drawn from identical populations. The Z-statistic was computed in the following way:

$$Z = \frac{\mu_f - \mu_{nf}}{\sqrt{\sigma_f^2/n_f + \sigma_{nf}^2/n_{nf}}}. \tag{9}$$

Here, $\mu_f$ and $\mu_{nf}$ correspond to the means of the Stokes $Q$, $U$ values and the polarization degree of the flat-fielded and unflat-fielded data, respectively, $\sigma_f$ and $\sigma_{nf}$ correspond to their standard deviations, and $n_f$ and $n_{nf}$ to the number of data points in each sample. From our data, $Z_{Q,V} = 1.52$, $Z_{U,V} = -1.35$, $Z_{P,V} = 0.58$, $Z_{Q,R} = 0.54$, $Z_{U,R} = 0.75$, and $Z_{P,R} = 1.45$. Setting $\alpha = 5\%$, we cannot reject the null hypothesis of the two samples being drawn from the same distribution with a 95% confidence level. As a consistency check, we carried out the same exercise, but rather than comparing two samples of different stars, we compared the derived polarimetric values of the same science frames, both applying and not applying the flat-field correction to them, thus allowing for a one-on-one comparison. As expected, a Z-test and a Kolmovorov–Smirnov test[51] revealed no significant difference between the two samples. We thus conclude that, within the precision of our data, we do not observe any significant effect introduced by our flat-fielding procedure.

### 4.6 Case of 1ES 1101–232

As an illustrative example, we show results on the blazar 1ES 1101-232 22, which was observed as part of our photo-polarimetric monitoring of blazars. This BL Lac object shows a changing polarimetric behavior with time. One of its first polarimetric measurements in the optical was given by Ref. 52, who reported a maximum polarization degree of 2.7%, while Ref. 53 observed the largest value in polarization degree ever reported so far (∼14.7%) for this object. Besides that strong change in polarization degree, the blazar is of our particular interest because it is relatively nearby ($z = 0.186$).[54] Hence, it shows a resolved galaxy that will allow us to test our method to correct the polarimetric data by the depolarizing effect introduced by the host galaxy and by the changes in seeing taking place during the observing runs. For an in-detailed description and motivation of these corrections, see Ref. 19.

We observed the blazar 1ES 1101-232 during OC-5 (March 14, 16, and 17, 2018) in the $R$ band. We collected a total of 16 polarimetric points with typical exposure times of 180 s. We corrected the data for instrumental polarization and foreground polarization following.[19] In addition, we transformed the polarization angles to the standard system using data from highly polarized standard stars. Figure 8 shows the time evolution of the polarimetric parameters of 1ES 1101-232 and field stars (FS-1, FS-2, FS-3, and FS-4), corresponding to March 16. 1ES 1101-232 shows a marginal evidence of intranight variability, manifested as a decrease and posterior increase of ∼1% in the degree of linear polarization. Regardless this low-amplitude variability, the low averaged polarization degree ($P = 1.39 \pm 0.27\%$) could be indicative of a current low activity state. The angle is steady, with a mean value $\langle \theta \rangle = 196.2 \pm 4.8$ deg. Figure 8 also shows the behavior with time of the polarimetric parameters of four field stars indicated with triangles. The mean polarization of the brightest field stars is roughly consistent with the expected interstellar polarization, $P_{IS} \leq 0.5\%$.[55] The faintest star (FS-3), in turn, shows a higher polarization degree, although in this case, the S/N ratio is poorer. In the same way as in Sec. 4.4, we tested the stability of the polarimetric parameters for the field stars. We found no significant serial correlation in the polarization values. Furthermore, Fig. 9 shows the behavior of the polarization as a function of the standard $R$ magnitude of the blazar and the field stars, collected on March 16. We observe a moderate increase of the individual polarimetric errors and the scatter of the polarimetric points for weaker magnitudes. This is expected to occur and directly linked to a detriment in the SNR of the photometry of weaker stars for a fixed exposure time. In the case of 1ES 1101-232, the observed scatter it could be associated with the possibility that the polarization of the blazar varies intrinsically, as shown in Fig. 8. Finally, we measured a mean standard magnitude in the $R$ band for 1ES 1101-232 of 16.1(3) mag. A complete study of this source will be shown in Sosa et al. 2019 (in preparation).

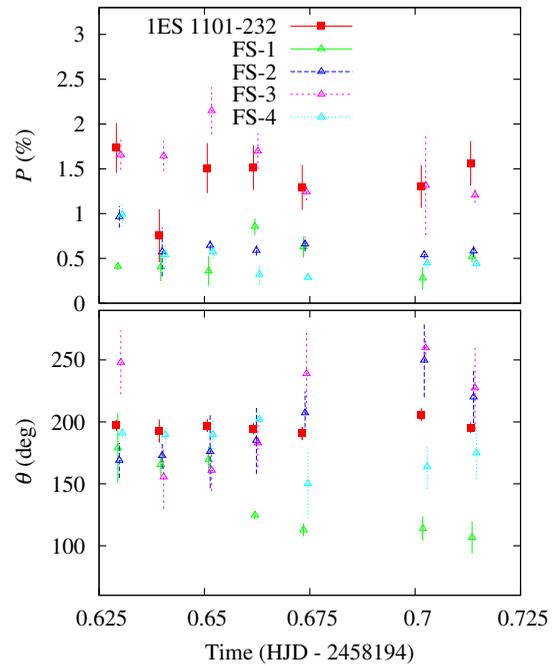

**Fig. 8** Time evolution of the polarimetric parameters for 1ES 1101-232 (red squares) and field stars (colored empty triangles). The data points correspond to the $R$ band and have been corrected by instrumental polarization. (a) The polarization degree in percentage. (b) The polarization angle in degrees.

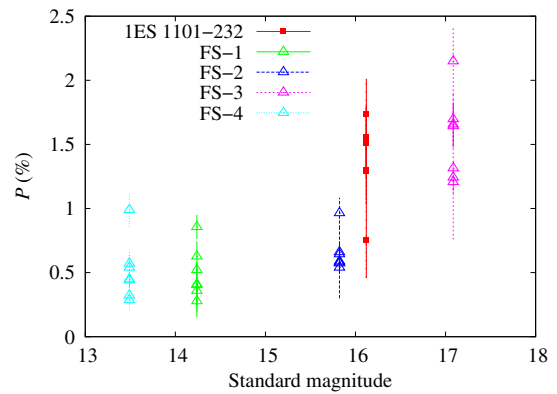

**Fig. 9** Polarization degree in percentage as a function of the standard magnitudes of the blazar and the field stars, FS, for the $R$ band. The symbols are equivalent to the ones in Fig. 8.

## 5 Conclusions

In this work, we studied the behavior of the instrumental polarization of CasPol, a dual-beam imaging polarimeter mounted at the Argentinean 2.15-m Jorge Sahade Telescope. For this end, we observed 12 polarized and unpolarized standard stars spread along five OCs during dark nights. After making a detailed analysis of the aperture size for optimum polarimetric measurements, we characterize the instrumental polarization of CasPol to be of ∼0.2% for the three bands. From the observation of unpolarized standard stars evenly distributed on the CCD, we estimated a negligible dependence of instrumental polarization on position for the $V$ and $R$ bands, and only a marginal dependence for the $I$ band. Our derived Stokes $U$, $Q$ parameters, (as well as their corresponding polarization degree and angle) for several polarized and unpolarized standard stars are compared to published values, when possible, showing consistency at a minimum of $2\sigma$ level. We made an in-depth comparison between our observed polarized standard star, Ve6 23, to the values reported by Ref. 26. In all cases, we find consistent results, with the exception of the $I$ band. We also report values for the polarization state of our observed standard stars. Furthermore, we determine that flat-fielding does not introduce any significant (within the precision of our data) instrumental effect to the resulting polarimetric states by comparing a large sample of unpolarized standard stars that were (and were not) calibrated with flat-fields. We determine the stability with the polarization angle to be ∼0.3 deg, and we do not find any significant dependency of this stability with wavelength. Overall, CasPol is a well-behaved optical dual-beam polarimeter, allowing researchers to carry out follow-up campaigns with reliable, stable measurements. With respect to the unpolarized standard star HD 64299, the values computed in this work in the $V$-band are consistent within $1\sigma$ uncertainties with Refs. 21 and 25. We believe that it is necessary to perform repeat observations of this star to establish its polarization state. Finally, we tested the instrument through observations of the blazar 1ES 1101-232. We analyzed the behavior of its linear polarization computing the parameters $P$ and $\theta$ corresponding to March 16. We measured $P = 1.39 \pm 0.27\%$, which would indicate that the blazar is currently in a low activity state, since values as high as $P \sim 14.7\%$ were reported in the literature.

### Acknowledgments

This work was funded with grants from Consejo Nacional de Investigaciones Científicas y Técnicas de la República Argentina and Universidad Nacional de La Plata (Argentina) and based on observations carried out at the Complejo Astronómico El Leoncito, operated under agreement between the Consejo Nacional de Investigaciones Científicas y Técnicas de la República Argentina and the National Universities of La Plata, Córdoba, and San Juan. Cv.E. acknowledges funding for the Stellar Astrophysics Centre, provided by The Danish National Research Foundation (Grant DNRF106). M.S. acknowledges the Stellar Astrophysics Centre for funding a fruitful visit. The authors acknowledge the staff of CASLEO for an outstanding support during the OCs.

**Marina Sosa** is currently in the final stages of her PhD studies at the Instituto de Astrofísica de La Plata, Argentina. Since 2011, she has been researching blazars, mainly studying their photopolarimetric light. Her main interests are high-energy blazars and the connection between their emission in the optical and high-energy wavelengths. Her contribution to the field accounts with a detailed method to correct the polarization of blazars for the effects introduced by the host galaxy.

**Carolina von Essen** is currently an assistant professor at the Stellar Astrophysics Centre, Denmark. Since 2010, she has been investigating exoplanets in great detail. Her main areas of research are the characterization of the chemical composition of exoplanet atmospheres, and the determination of exoplanetary masses through dynamical interactions. Her contributions to the field include, among others, the modeling of exoatmospheric absorption features and the first discovery of aluminum oxide in an ultra-hot Jupiter.

**Ileana Andruchow** received her PhD in astronomy. She is a researcher for the CONICET (Argentinan Scientific Council) and a teacher at La Plata University. Her main interests are photometry variability of blazars and optical linear polarization behavior analysis of AGNs, particularly, blazars. She also worked with optical interferometry, using VLTI data to analyze the structure of B[e] supergiant stars.

**Sergio A. Cellone** is presently an associate professor at La Plata National University, independent researcher (CONICET), and director of El Leoncito Astronomical Complex (CASLEO), Argentina. He has been the head of the Argentinean National Gemini Office. His main research interests deal with observational astrophysics, involving blazars and dwarf galaxies, using different optical techniques (imaging, low-resolution spectroscopy, and polarimetry).

**Luis A. Mammana** is currently support astronomer and vice director of the Complejo Astronómico El Leoncito (CASLEO, Argentina). He received PhD in astronomy from the Universidad Nacional de La Plata, Argentina, and his field of work is mainly planetary sciences. He studies populations of centaurs, comets, transit of extrasolar planets, etc. He has also participated in studies of blazares. In recent years, he has worked alongside specilized engineers in astronomical instruments.